\documentstyle[12pt,twoside,fleqn,epsfig,psfig,espcrc1]{article}
\title{The $u-d$ quark mass difference and nuclear charge symmetry
breaking}
\author{Sidney A. Coon\address{Institute for Nuclear Theory, University
of Washington, Seattle, WA 98195 and
Physics Department, New Mexico State
University, Las Cruces, NM 88003, USA}}

\begin{document}
\maketitle

\begin{abstract}
		
The group theoretical analysis of the Coleman-Glashow tadpole  picture
of ``meson-mixing" is quantitatively reproduced by the $u-d$
constituent quark mass difference in quantum loop calculations of the
self-energies of mesons. This demonstration that the Coleman-Glashow
scales can be directly calculated from the constituent $u - d$ quark
mass difference finishes the link between charge symmetry breaking in
low energy nuclear physics and the $u - d$ quark mass difference in
particle physics.		
\end{abstract}

\section{INTRODUCTION}

The very recent precision measurement of the $\Xi^0$
mass~\cite{ximass} has revived interest in the electromagnetic {\em
(em)} mass splittings of the baryons, because of the resistance of
the forty-year old Coleman-Glashow relation \cite{CG1} to substantial
symmetry-breaking effects in quark masses.  This relation
\begin{equation}
	M_{\Xi^-} - M_{\Xi^0} = M_{\Sigma^-} - M_{\Sigma^+} + M_p -
M_n  \, , \label{CGrel}
\end{equation}
was derived assuming unbroken flavor $SU(3)$ and is now satisfied to
within $4\pm3\%$ (scaled by $M_{\Sigma^+} - M_{\Sigma^-} \sim 8$ MeV),
or broken at the one sigma level depending on your point of
view~\cite{others}.  While eq. (\ref{CGrel}) was derived from only the
isospin breaking electromagnetic interaction,
the individual $\Delta I =1$ baryon pairs should, however, reflect
$SU(2)$ breaking caused by $\Delta I =1$ quark mass differences. 
Subsequent to Ref.~\cite{CG1} (and before the quark picture), Coleman  and
Glashow~\cite{CG} suggested that symmetry-violating processes are
dominated by symmetry-breaking tadpole diagrams with scalar mesons
linking the tadpole to the $SU(3)$ invariant strong interactions.  To
describe  electromagnetic splittings, they combined the tadpole
Hamiltonian $ H^3_{tad}$ (transforming like $\lambda_3$ in an $SU(3)$
context) together with the current-current operator $H_{JJ}$
(corresponding to the first order  in $\alpha$ contribution due to
photon exchange) to form an effective $\Delta I =1$  electromagnetic
{\em (em)} Hamiltonian density operator,
\begin{equation}
 H_{em} = H_{JJ} + H^3_{tad}\;.  \label{hem}
\end{equation}
In a group-theoretical sense, eq.~(\ref{hem}) gives a universal
$\Delta I =1$ 
picture~\cite{CS} of
\begin{itemize}
	\item[(a)] hadron electromagnetic mass splittings of
	pseudoscalar (P) and vector (V) mesons, along with the
splittings
	of octet baryons (B) and decuplet (D) baryons,
	\item[(b)] off-diagonal {\em em} transitions 
	$\langle\Sigma^\circ|H_{em}|\Lambda\rangle$,
$\langle\pi^\circ|H_{em}|\eta_{NS}\rangle$, and 
$\langle\rho^\circ|H_{em}|\omega\rangle$,
\end{itemize}
where $\eta_{NS}$ is the non-strange $\bar{q}q$ component of the
$\eta$
and the $\omega$ is 97\% nonstrange.

Quite soon after Ref.~\cite{CG}, Dalitz and von Hippel~\cite{DvH} applied the
Coleman-Glashow (CG) Hamiltonian operator  to the issue of charge
symmetry for the $\Lambda$ hyperon and, in particular the charge
asymmetry of the $\Lambda N$ interaction (i.e.~$\Lambda n$ versus
$\Lambda p$).  The CG operator, illustrated in Figure 2 for this case,
 suggests an appreciable $\Delta I =1$
transition between the isospin-pure $\Sigma^\circ$ and $\Lambda$
hyperons (often referred to as electromagnetic mixing). 
\begin{figure}[htbp]
\unitlength1.cm
\begin{picture}(6,2.5)(2,17)
\includegraphics{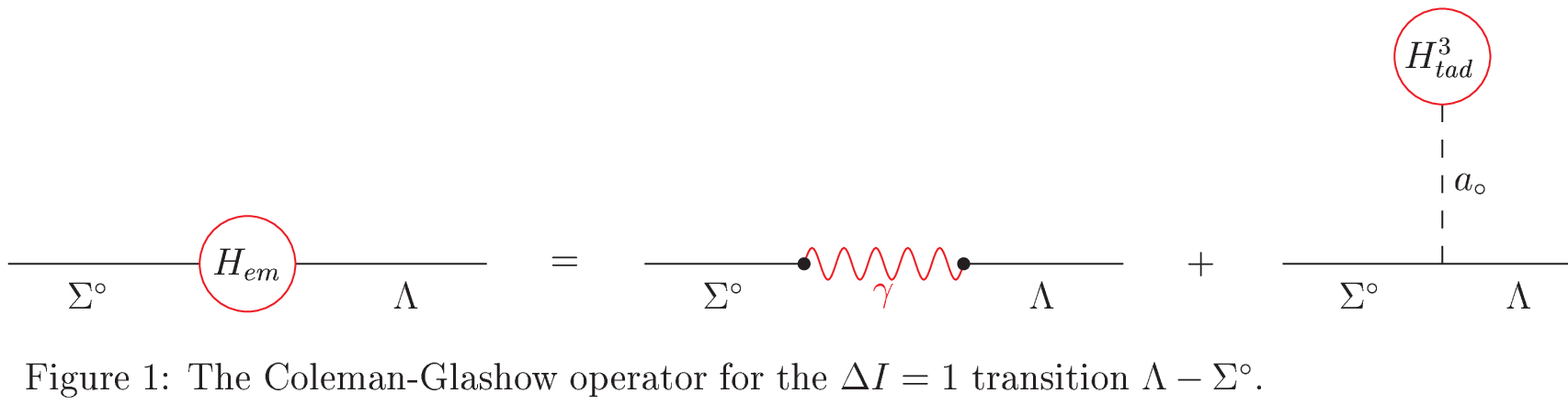}
\end{picture}
\end{figure}
This transition allows an exchange of the isospin one pion between the
$\Lambda$ and nucleon, otherwise forbidden by isospin conservation,
which contributes to hypernuclear charge symmetry breaking.
\begin{figure}[htbp]
\unitlength1.cm
\begin{picture}(6,3.5)(1,14)
\includegraphics{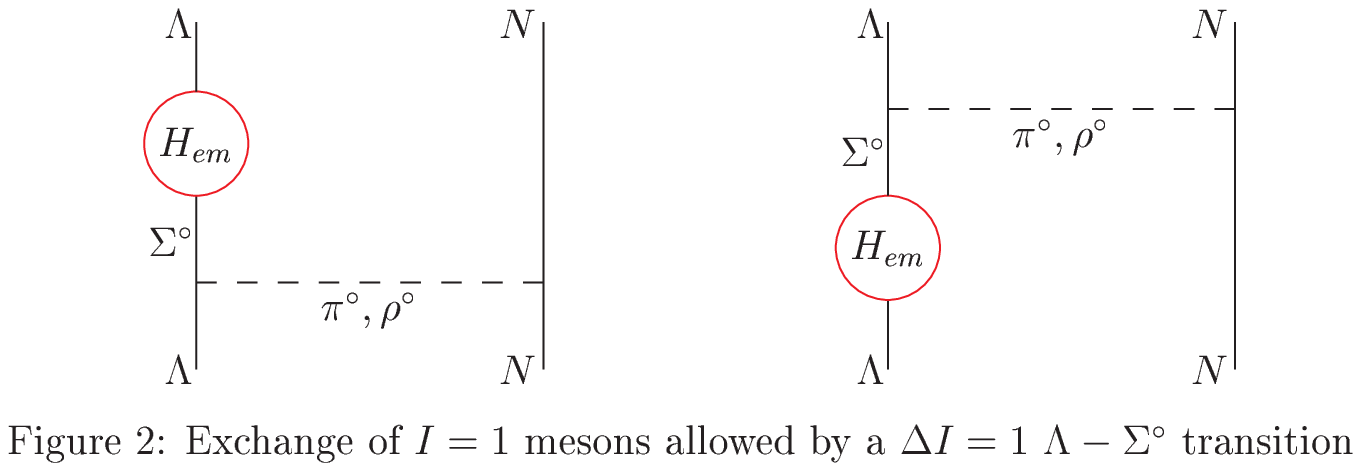}
\end{picture}
\end{figure}
Unfortunately our experimental knowledge of the $\Lambda N$
interaction has not progressed much in the last thirty six years
\cite{Gibbs}, so the success of the Coleman-Glashow   off-diagonal
{\em em} transitions $\langle\Sigma^\circ|H_{em}|\Lambda\rangle$, 
$\langle\pi^\circ|H_{em}|\eta_{NS}\rangle$,  and 
$\langle\rho^\circ|H_{em}|\omega\rangle$  in describing hypernuclear
charge symmetry breaking remains problematic at
present~\cite{pruhonice}.

The experimental knowledge of charge asymmetry in the $NN$ interaction,
on the other hand, is rather good and the  off-diagonal {\em em}
transitions,  $\langle\pi^\circ|H_{em}|\eta_{NS}\rangle$ and 
$\langle\rho^\circ|H_{em}|\omega\rangle$, from the Coleman-Glashow {\em
em} Hamiltonian operator embedded in single meson exchange (tree)
diagrams analogous to those of Fig. 2 give the dominant and
satisfactory description of nuclear charge asymmetry~\cite{Dubna}. But
it is the group theory structure and single universal Coleman-Glashow
scale  which ``does the heavy lifting" of this description (although
the measured properties of their postulated $I = 1$ scalar meson, now
called the $a_0$, do  recover  the tadpole scale~\cite{CS98}).  Indeed,
Coleman and Glashow emphasized that ``our explanation of
symmetry-breaking phenomena suggests, but does not require, the
existence of scalar mesons"~\cite{CG}. It is the purpose of this talk
to  show that the group theoretical analysis of the CG tadpole
(reviewed in more detail in~\cite{CS}) is quantitatively reproduced by
the $u-d$ constituent quark mass difference in quantum loop
calculations of the self-energies of mesons.  But first I review the
tadpole scale and show how the scale of the baryons is the same as the
tadpole scale established by electromagnetic meson mass splittings.

\section{MASS SPLITTINGS AND THE TADPOLE SCALE}

 Returning to electromagnetic mass splitting of the baryons,  the new
measurement \cite{ximass} is within two standard deviations of the the
earlier CG tadpole prediction~\cite{CS} of eq.~(\ref{hem}) for the
octet baryons, reproduced  in Table I. In this Table, the self-energy
shift of a hadronic state arising from single photon exchange
($H_{JJ}$) has been evaluated via  the known dominance of the Born
terms in the dispersive evaluation of the Cottingham formula. I show
two sets of results for the octet baryons, and note that an early
evaluation (including estimates of smaller resonance and other
contributions) in 1969 \cite{Bucella} is  strikingly confirmed by a
recent (1997) lattice QCD calculation which summed the electric and
magnetic Born contributions over discretized bosonic momenta of the
finite lattice \cite{lattice}. To estimate the universal value of the
tadpole from the baryon octet, one should concentrate on line three
where only the small magnetic Born terms contribute to $H_{JJ}$ of
$m_{\Sigma^+}-m_{\Sigma^-}$ and  $H^3_{tad}$ is isolated.  Then one
fills out Table 1 with
\begin{equation}
(H^3_{tad})_{p-n } =  \frac{2}{3}(H^3_{tad})_{\Sigma^{\circ} - \Sigma^-}
    = \frac{1}{3}(H^3_{tad})_{\Sigma^{+} - \Sigma^-}
    = \frac{1}{2}(H^3_{tad})_{\Xi^{\circ} - \Xi^-} \approx -2.5 \,\,
{\rm MeV}\; , \label{octet}
\end{equation}
assuming the same semi-strong and electromagnetic $d/f$ ratio of
$-1/3$  (see eg. Ref.~\cite{CS}), and obtains the CG tadpole
prediction~\cite{CS} of eq.~(\ref{hem}) for the octet baryons. 
\vspace{-.45in}
\begin{table}[htbp]
\begin{center}
 \begin{minipage}{14cm}
  \renewcommand{\footnoterule}{}
  \begin{center}
	\caption{$SU(2)$ mass splitting for octet baryons (in
         MeV)}\label{minitab1} 
	\vspace{.1in}
        \begin{tabular}{crrrrrc} \hline
         Baryons & $H_{JJ}$\footnote{Cottingham formula evaluated in
	 Ref.~\cite{Bucella}} & 
        $H_{JJ}$\footnote{Cottingham formula evaluated in Ref.~\cite{lattice}}
	 & $H^3_{tad}$ & 
	Total(a) & Total(b)
        & Experiment\footnote{Particle Data Group \cite{PDG}}  \\  \hline
        $m_p-m_n$ & +0.8 & +0.75 &   -2.5 & -1.7 & -1.75 & -1.29  \\
        $m_{\Sigma^0}-m_{\Sigma^-}$ & -1.0 &-0.86 & -3.75 & -4.75 &
        -4.61 &-4.81$\pm$0.04  \\
        $m_{\Sigma^+}-m_{\Sigma^-}$ & -0.3 & -0.06 & -7.5 & -7.8 &
         -7.56 & -8.08$\pm$0.08  \\
        $m_{\Xi^0}-m_{\Xi^-}$  & -1.1 & -0.86 & -5.0 & -6.1 &
          -5.9 &-6.50$\pm$0.25\footnote{new measurement \cite{ximass}}  \\ 
        \hline 
       \end{tabular}
       \vspace{-.1in}
    \end{center}
   \end{minipage}
  \end{center}
\end{table}
\vspace{-.45in}
 One could attempt to improve the description of $m_p-m_n$ by quoting
the  standard QED current-current (ie. photon exchange) self-energy for
charged protons 
\begin{equation}
 (H_{JJ})_p \simeq \Sigma(m) = \frac{3\alpha m_p}{2\pi}\left[
 \ln(\frac{\Lambda}{m_p}) + \frac{1}{4}\right] \simeq 1.2 {\rm~MeV},
\label{photon}
\end{equation}
and choosing the value of the ultraviolet cutoff $\Lambda = 1.05$ GeV 
so that (\ref{hem}) is compatible with the observed proton-neutron mass
difference of -1.29 MeV \cite{DLS}.   From  an effective field theory
viewpoint, the discrepancy between the $g$-factor of the relativistic
point spin $\frac{1}{2}$ proton and the measured $g$ of the proton
suggests that  the cutoff in nucleon QED must be of order of the
nucleon's mass $m_p$~\cite{Lepage}, consistent with (\ref{photon}).
This cutoff exercise (which neglects magnetic moment contributions to
the fermion's self-energy) yields an $H_{JJ}$ somewhat larger than  the
Born contributions of Table I and suggests that one should not limit
oneself to  estimates of $H_{JJ}$ and $m_d - m_u$ made only from the
non-strange sector~\cite{Bira}, ie. $m_p-m_n$ and electromagnetic pion
mass differences, the latter to which we now turn.

The Coleman-Glashow postulate (1) combined with the Dashen PCAC
observation~\cite{RD} that, in the soft limit, 
neutral pseudoscalar meson $H_{JJ}$ matrix elements vanish
\begin{math}
(H_{JJ})_{\pi^\circ} = (H_{JJ})_{K^\circ} =
(H_{JJ})_{\bar{K}^\circ} = (H_{JJ})_{\pi^\circ\eta} = 0  
\end{math}
but charged matrix elements 
\begin{math}
(H_{JJ})_{\pi ^+} = (H_{JJ})_{K ^+}  
\end{math}
do not vanish, can be related by group theory to the measured em mass splittings
\begin{equation}
\Delta m^2_K 
\equiv m^2_{K^+}-m^2_{K^\circ}\approx -3960\; {\rm MeV}^2, \quad
\Delta m^2_\pi\approx 1260\; {\rm MeV}^2\;\;.  \label{del_m}
\end{equation}
This predicts
\begin{equation}
(H_{em})_{\Delta \pi}=(H_{JJ})_{\Delta \pi} + (H^3_{tad})_{\Delta \pi} 
  =  (H_{JJ})_{\Delta K} + 0 = \Delta m^2_\pi\;, \label{eqno4a}
\end{equation}
\begin{equation}(H_{em})_{\Delta K}=(H_{JJ})_{\Delta K} + (H^3_{tad})_{\Delta K} 
  = \Delta m^2_K \;,   \label{eqno4b}
\end{equation}
and subtracting (\ref{eqno4a}) from (\ref{eqno4b}) fixes the 
kaon tadpole scale~\cite{CS}
\begin{equation}
(H^3_{tad})_{\Delta K} = \Delta m^2_K    -
\Delta m^2_\pi\approx-5220\:  {\rm MeV}^2\; .  \label{ktad}
\end{equation}
A number of model and lattice calculations suggest that for physical 
pseudoscalar mesons 
$(H_{JJ})_{\Delta \pi}\approx \Delta m^2_\pi\approx 1260\; {\rm MeV}^2$,
but that $(H_{JJ})_{\Delta K}\approx 1900-2600\; {\rm
MeV}^2$\cite{K_JJ}.  If so, then the kaon tadpole scale increases
slightly from (\ref{ktad}) to  $-5800$ --- $-6500$ MeV$^2$.  
In section 3, we
will show how to recapture the soft kaon tadpole scale of (\ref{ktad})
and Ref. \cite{CS}.
Also $SU(3)$ symmetry predicts the off-diagonal $\Delta I =1$
transitions as~\cite{CS,angle}
\begin{eqnarray}
\langle\pi^\circ|H^3_{tad}|\eta_{NS}\rangle &=& 
\Delta m^2_K - \Delta m^2_{\pi}\approx -5220\; {\rm MeV}^2 \;,
\label{pieta}\\ 
\langle\rho^\circ|H^3_{tad}|\omega\rangle &=&
\Delta m^2_{K^*} -\Delta  m^2_{\rho}\approx -5130\; {\rm MeV}^2 \;.
\label{rhoomega}
\end{eqnarray}
The meson scale of about $-5200$ MeV$^2$ in (\ref{ktad}),(\ref{pieta}), and
(\ref{rhoomega}), extendable to vector mesons via $SU(6)$ \cite{CS}
or by the measured properties of the $a_\circ$ \cite{CS98}, can be
related to the fitted baryon scale (\ref{octet}) by multiplying 
the latter by the normalization of the baryon spinors: $\bar{u}
u=2M_{\rm baryon} \approx 2300$ MeV. Then  the mass$^2$ version of 
(\ref{octet}) is $(H^3_{tad})_{B}\approx -5700$ MeV$^2$ indicating a
universal Coleman-Glashow scale and a picture of electromagnetic mass
splittings which almost certainly
rests upon the up-down quark mass difference \cite{Millerrev3}.

\section{QUARK LOOPS AND UP-DOWN CONSTITUENT QUARK MASS DIFFERENCE}
A   constituent quark mass difference 
\begin{math}
	m_d - m_u \approx 4\:  {\rm MeV}
\end{math}
follows from the observed  $(\bar{s}d)K^0 - (\bar{s}u)K^+$ mass
difference:
\begin{math}
   m_{K^\circ}(497.67) - m_{K^+}(493.68) \simeq m_d - m_u 
   = 3.99\:  {\rm MeV}\; ,  \label{Kdiff}
\end{math}
with the common $\bar{s}$ spectator quark subtracting out of the kaon
mass difference.  Also the $\Sigma$ baryon mass difference in quark
language for $(sdd)\Sigma^- - (suu)\Sigma^+$ is
\begin{math}
m_{\Sigma^-}(1197.45) - m_{\Sigma^+}(1189.37) \simeq 2(m_d - m_u) 
   = 8.08\:  {\rm MeV}\; ,  \label{Sdiff}
\end{math}
with common $s$ spectator mass subtracting out and the photon
interaction $H_{JJ}$ at a minimum (Table I). These estimates are
consistent and hint that a consistent Coleman-Glashow picture of
hadronic mass splitting and mixings could be obtained from the {\em
differences} of constituent quark loop diagrams~\cite{DLS,SC00}.

Returning to the kaon tadpole scale of $-5220$ MeV$^2$ in eq.
(\ref{ktad}), in quark language this is due to the quark line graphs of
Figure 3.  They are the $u-d$ kaon quark bubble graphs plus the
difference of   those $u-d$ quark loops which look like a $\Delta I =
1$ $a_\circ$ tadpole with $a_{\circ} KK$ coupling~\cite{DLS,SC00}. 
\begin{figure}[htbp]
\unitlength1.cm
\begin{picture}(6,6.25)(0,11.25)
\includegraphics{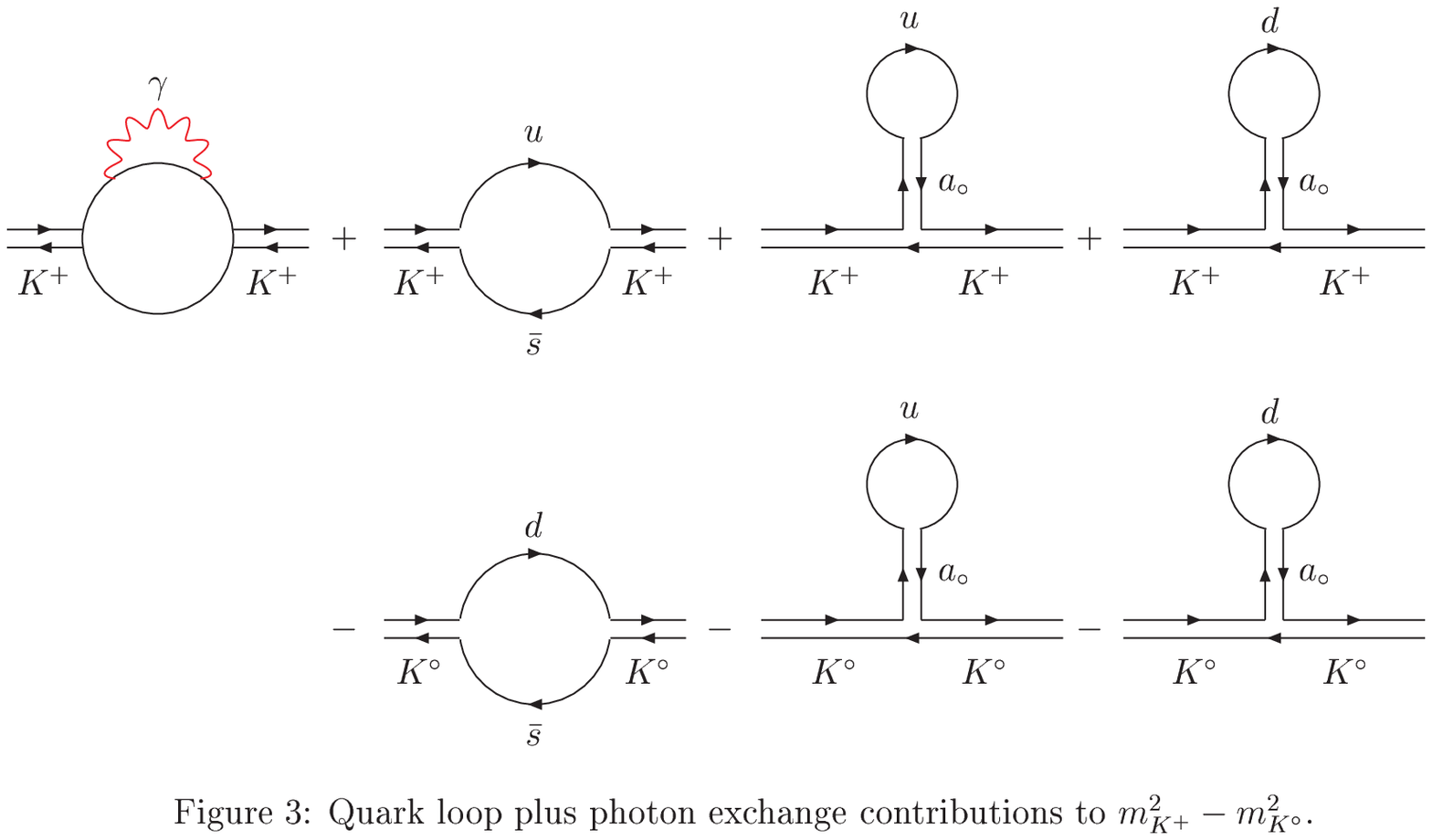}
\end{picture}
\end{figure}
Evaluation of these
quark loop graphs results in the soft momentum limit to
\begin{eqnarray}
     (\Delta m ^2_K)_{qk\: loops} &=& (m_u - m_d) [2(2\hat{m} - m_s) 
     + 6\hat{m}^2(m_s -\hat{m})/(2\hat{m}^2 + m_s^2)] \nonumber \\
    &  & \mbox{} +  8 g_{a_\circ KK}(m_u - m_d) \hat{m}^2/g m^2_{a_0}  \;,     
        \label{kloops}
\end{eqnarray}
where $\hat{m} = (m_d + m_u)/2 \approx 337$ MeV, and $m_s/\hat{m}
\approx 1.44$. Each of the two pairs of $u-d$ tadpole-like graphs of Figure
3, summed in the second line of (\ref{kloops}),  includes the integral
($\bar{d}^4p \equiv d^4p/(2\pi)^4$)
\begin{equation}
  I = \int \bar{d}^4p \left[
      \frac{m_u}{p^2-m_u^2} - \frac{m_d}{p^2-m_d^2}\right]
= i(m_u - m_d)\frac{\hat{m}^2}{N_c g^2}\; , \label{identity}
\end{equation}
with $N_c g^2 = 4\pi^2$~\cite{LSM}.  The value of this $u-d$ 
difference loop integral $I$, which looks like the  head of a tadpole and
appears repeatedly in the following, is  independent of the
regularization scheme used to obtain it~\cite{DSR}.
To evaluate the four tadpole graphs summed in the second line of
(\ref{kloops}), we   use the  linear  $\sigma$
model Lagrangian  quark-quark-meson coupling  $g = 2\pi / \sqrt{3} \approx
3.63$~\cite{LSM} but deviate somewhat from the linear  $\sigma$ model Lagrangian
tri-meson coupling $g_{a_\circ KK} = (m^2_{a_\circ} - m^2_K)/(2f_K)\approx
3.15$ GeV, and use instead $g_{a_\circ KK}\approx 2.7$ GeV, the latter
value an average of this chiral symmetry estimate and the $U(3)$
symmetry estimate of $g_{a_0KK} = g_{\sigma \pi \pi}/2 = m^2_\sigma/ 2
f_\pi \approx 2.55$ GeV~\cite{DLS}. 
Given the $d-u$ quark mass difference of about 4 MeV, eq.~(\ref{kloops})
leads to~\cite{SC00}
\begin{equation}
   (\Delta m^2_K)_{qk\: loops} \approx -(2384 + 2800)\:  {\rm MeV}^2
   \approx -5184 \:  {\rm MeV}^2\; . 
   \label{kloops_num}
\end{equation}
Note that eq.~(\ref{kloops_num}) is in agreement with the soft
Coleman-Glashow $\lambda^3$ kaon tadpole in (\ref{ktad}) as found in
Ref.~\cite{CS}.

Next we compute the nonstrange (NS) $\Delta I = 1$ {\em em} transition
amplitude $\langle\pi^\circ|H^3_{tad}|\eta_{NS}\rangle$ for
$m_{\eta_{NS}} \approx 760$ MeV (the weighted average of the
$\eta(548)$ and  $\eta'(958)$\cite{angle}) in terms of   $u-d$  quark
bubble {\em and} $u-d$ $a_{\circ} $ tadpole graphs analogous to those
of Figure 4.  These graphs give~\cite{SC00}

\begin{eqnarray}
(H_{\pi \eta_{NS}})_{qk \:loops} &= & 
 (m_u - m_d)(2\hat{m} +16 \hat{m} ^3/ m^2_{a_\circ}) \nonumber \\
   & \approx & -2696\:{\rm MeV}^2 - 2535\:{\rm MeV}^2 \quad
   \approx \;\;  -5231\:{\rm MeV}^2\;,
				\label{pieta_q}
\end{eqnarray}
where the $(m_u - m_d)2\hat{m}$ factor in (\ref{pieta_q}) derives from
the difference of $u-d$ bubble graphs:
\begin{equation}
-4ig^2N_c\int \bar{d}^4p\left[
     \frac{1}{p^2-m_u^2}-\frac{1}{p^2-m_d^2}\right]\;. \label{bubdiff}
\end{equation}
The integrand of (\ref{bubdiff}) can immediately be turned into $(m^2_u-
m^2_d)/(p^2-m_q^2)^2$ so that the $u-d$ bubble graphs expression
(\ref{bubdiff}) is simply $(m^2_u-m^2_d)$ times 
the logarithmic divergent gap equation 
\begin{equation}
1 =-4iN_cg^2\int \frac{\bar{d}^4p}{(p^2-m_u^2)(p^2-m_d^2)},
\end{equation}
which arises from the quark loop generation of the the decay constant
$f_\pi$ in the quark-level Goldberger-Treiman relation $f_\pi g =
\hat{m}$~\cite{LSM,DSR}.
The $16 \hat{m} ^3/ m^2_{a_\circ}$ term in (\ref{pieta_q}) stems from
the $u-d$ quark loop which looks like a $\Delta I = 1$ $a_\circ$
tadpole with $a_{\circ}\pi \eta_{NS}$ coupling.  This latter $u-d$
quark loop {\em difference} integral $I$  is again evaluated via equation
(\ref{identity}) above.  To obtain the number (and the cubic power of
$\hat{m}$) in (\ref{pieta_q}) we have replaced the $a_{\circ}\pi
\eta_{NS}$ coupling  $(m^2_{a_\circ} - m^2_{\eta_{NS}}) / 2 f_\pi$ by
its Lagrangian linear $\sigma$ model analogue~\cite{LSM} $
(m^2_{\sigma} - m^2_\pi)/2f_\pi = (4 \hat{m})^2/2 f_\pi$.  Then $f_\pi
\approx 93$ MeV and  $\hat{m} \approx 337$ MeV, so that the
Goldberger-Treiman relation  $f_\pi g =\hat{m}$ remains valid (for $g =
2\pi/\sqrt{3}$~\cite{LSM}).

\begin{figure}[htbp]
\unitlength1.cm
\begin{picture}(6,2.7)(0,16.5)
\includegraphics{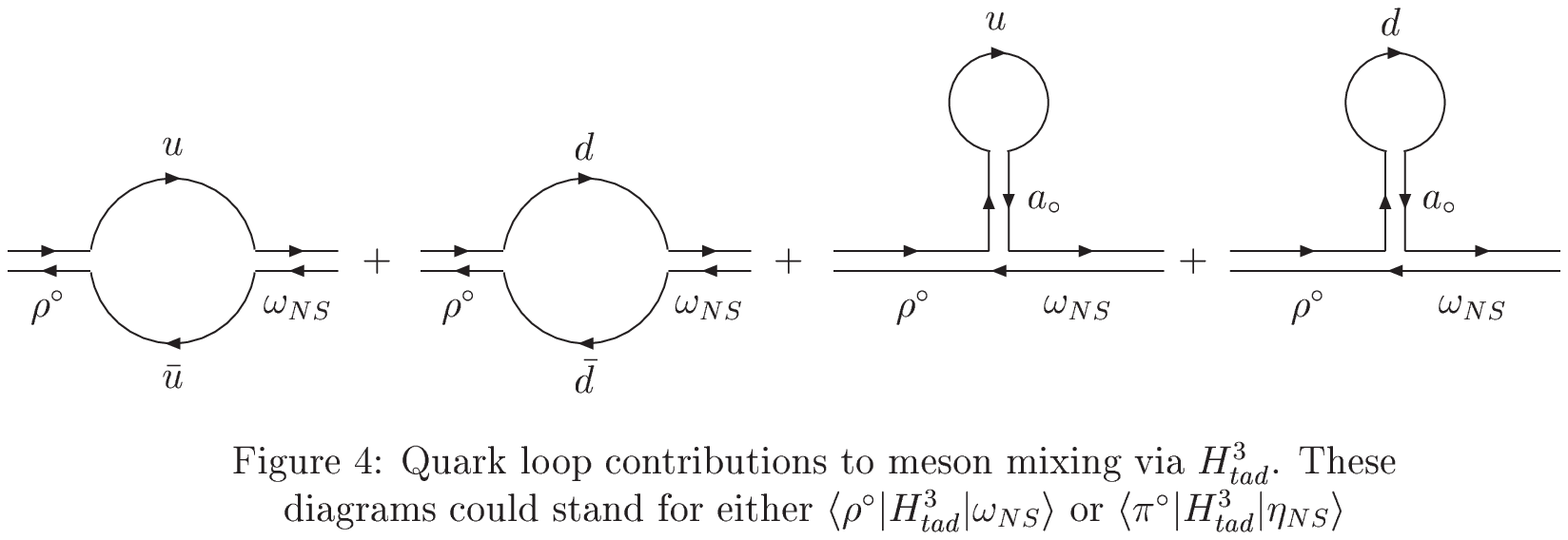}
\end{picture}
\end{figure}

Lastly we study the $\Delta I = 1$ {\em em} transition amplitude
$\langle\rho^\circ|H^3_{tad}|\omega_{NS} \rangle$ in terms of the
$u-d$  quark diagrams of Figure 4.   For this vector meson
transition one must work with the QED-like nonstrange quark bubble
polarization tensor~\cite{Del}, taking the $SU(3)$ value $g_{\omega} = 3
g_{\rho}$, 
\begin{equation}
 \Pi_{\mu\nu}=(-k^2 g_{\mu\nu} +k_\mu k_\nu)\Pi(k^2,m_q^2) 
 g_\rho g_\omega / 12\;\;,  \label{Pimunu}
\end{equation}
where $g_{\rho}\approx 5.03$ and $g_{\omega} \approx 17.05$ (for 
$e=\sqrt{4\pi\alpha} \approx 0.30282$) follow from 
electron-positron decay rates~\cite{CS98,PDG}. The $\omega$ is 97\%
nonstrange and 3\% strange so the $\Delta I = 1$  transition 
between the physical particles is well described by the difference of
the $u-d$ polarization
function  
\begin{equation}
 \Pi(k^2,m_u^2)-\Pi(k^2,m_d^2) =
    -\frac{N_c(m_u^2-m_d^2)}{2\pi^2k^2}  \label{Pi}
\end{equation}
where $N_c$ is the number of colors in the quark model.  We emphasize
that the difference between the $u$ and $d$ quark contributions to the
polarization function is finite. In order for the
inverse propagator 
$\Delta^{-1}_{\mu\nu}(k)=(k_\mu k_\nu
 - k^2 g_{\mu\nu})[1 +   \Pi(k^2)] =-g_{\mu\nu}(k^2-m^2) +$ terms in $k_\mu
k_\nu$ to actually simulate a reciprocal vector meson propagator, 
it is clear that the polarization function $-k^2\Pi(k^2)$
in (\ref{Pi}) acts as  a squared (quark) mass.  Note that this
discussion and eq.~(\ref{Pi}) allows us to interpret the (squared mass)
bubble Hamiltonian density $\rho-\omega$ transition matrix element as 
\begin{eqnarray}
(H_{\rho \omega})_{qk \:loops}^{bubble} &=& -k^2[\Pi(k^2,m_u) -
\Pi(k^2,m_d)] g_\rho g_\omega / 12 \nonumber \\
  & = &  g_\rho^2 N_c(m_u^2-m_d^2)/8\pi^2  \; .   \label{rwbub}
\end{eqnarray}
Besides this $u-d$  quark
bubble term we must add in the  $\Delta I = 1$ $a_\circ$ tadpole term. 
It was shown in Ref.~\cite{CS98} that the measured decays of the $a_\circ$
meson with the aid of the vector dominance model lead to 
$g_{a_\circ \rho \omega} = 
g_{a_\circ \pi \eta_{NS}}$. Then the tadpole term for the $\rho\omega$
transition is the same as the second term of (\ref{pieta_q}), and 
the $\rho^\circ-\omega$ effective Hamiltonian becomes
\begin{eqnarray}
 (H_{\rho\omega})_{qk \:loops} &=& (m_u-m_d)[g^2_\rho N_c \hat{m}/4\pi^2
              + 16\hat{m}^3/m_{a_\circ}^2] \nonumber \\
	     & \approx& -2592\:{\rm MeV}^2
	      -2535\:{\rm MeV}^2 \quad
	      \approx \;\; -5127\:{\rm MeV}^2
	      \label{omrho_q}
\end{eqnarray}
for $N_c=3$,  $m_d - m_u \approx 4\:  {\rm MeV}$, and $\hat{m}\approx
337$ MeV~\cite{DLS}.
To compare with experiment, we must include the small
current-current photon exchange term~\cite{Gatto}:
\begin{equation}
	\langle\rho^\circ|H_{JJ}|\omega\rangle = 
	(e/g_{\rho})(e/g_{\omega})m^2_V
  \approx 644\;{\rm MeV}^2  \label{hjjvec}
\end{equation}
on the vector meson mass shell $k^2 = m_V^2$ in the spirit of vector
meson dominance (VMD). In (\ref{hjjvec}) we have used the average
$\rho^\circ-\omega$ mass  $m_V=776$ MeV  along with the updated VMD
ratios $g_{\rho}/e\approx16.6$ and  $g_{\omega}/e\approx56.3$, with the
latter $g_{\rho}$ and $g_{\omega}$  couplings found from
electron-positron decay rates~\cite{CS98,PDG}.  Combining (\ref{omrho_q})
with the $H_{JJ}$ term in (\ref{hjjvec}) according to the 
Coleman-Glashow decomposition (\ref{hem}) requires 
\begin{equation}
	(H_{em})_{\rho\omega} \approx (644 - 5127)\:{\rm MeV}^2
	\approx - 4483 \:{\rm MeV}^2\;\;.
\end{equation}
This latter scale derived from quark loops and photon exchange is quite
near the empirical $\Delta I = 1$ {\em em} transition $-4520 \pm 50
\:{\rm MeV}^2$ found from the measured
$\omega\rightarrow \rho^\circ \rightarrow 2 \pi$ decay
rate~\cite{CB,CS98}.

To conclude, the Coleman-Glashow group-theoretical decomposition
(\ref{hem}) leads to the {\em universal} $H^3_{tad}$ $\Delta I = 1$
scale of $\approx - 5200 \:{\rm MeV}^2$ in 
eqs.~(\ref{ktad},\ref{pieta},\ref{rhoomega}) which is close to the fitted
baryon tadpole scale $(H^3_{tad})_{p-n } \approx -2.5\:{\rm MeV}$ in
(\ref{octet}).  Both of these latter scales are reproduced in the
alternative quark-loop scheme, and
again result in the universal quark-loop  $\Delta I = 1$
transitions in eqs.~(\ref{kloops_num},\ref{pieta_q},\ref{omrho_q}),
which are based on $m_d - m_u \approx 4\:  {\rm MeV}$.

\end{document}